\documentclass[slac_one]{revtex4}
\usepackage{graphicx}
\usepackage{fancyhdr}
\newcommand{\Bot}{$b$}
\newcommand{\ttbar}{$t\bar{t}$}
\newcommand{\ET}{$E_T$}

\newcommand{\ipb}{$pb^{-1}$}
\newcommand{\lumi}{$1.9 fb^{-1}$}
\newcommand{\lxyCv}{176.7 }
\newcommand{\lxyUpStatErr}{10.0 }
\newcommand{\lxyDownStatErr}{8.9 }
\newcommand{\lxySystErr}{ 3.4 }
\newcommand{\lepPtCv}{173.5 }
\newcommand{\lepPtUpStatErr}{8.9 }
\newcommand{\lepPtDownStatErr}{9.1 }
\newcommand{\lepPtSystErr}{ 4.2 }
\newcommand{\combCv}{175.3 }
\newcommand{\combStatErr}{6.2 }
\newcommand{\combSystErr}{ 3.0 }
\newcommand{\lxyRes}{$ (\lxyCv ^{+\lxyUpStatErr}_{-\lxyDownStatErr} (stat) \pm \lxySystErr (syst))\ GeV / c^2 $ }
\newcommand{\lepPtRes}{$ (\lepPtCv ^{+\lepPtUpStatErr}_{- \lepPtDownStatErr} (stat) \pm \lepPtSystErr (syst))\ GeV / c^2 $}
\newcommand{\combRes}{$ (\combCv \pm \combStatErr (stat) \pm \combSystErr (syst))\ GeV / c^2$}
\newcommand{\lxyStatRes}{$ \lxyCv ^{+\lxyUpStatErr}_{-\lxyDownStatErr}\  GeV / c^2 $ }
\newcommand{\lepPtStatRes}{$ \lepPtCv ^{+\lepPtUpStatErr}_{- \lepPtDownStatErr}\  GeV / c^2 $}
\newcommand{\combStatRes}{ $ (\combCv \pm \combStatErr (stat))\ GeV / c^2 $}

\newcommand{\dataLepPt}{$55.2 \pm 1.3\ GeV / c$}
\newcommand{\dataLxy}{$0.596 \pm 0.017\ cm$}

\begin{document}

\title{{\small{Hadron Collider Physics Symposium (HCP2008),
Galena, Illinois, USA}}\\ 
\vspace{12pt}
Measurement of the Top Quark Mass using Quantities with Minimal Dependence on the Jet Energy Scale} 

%

\author{F. Garberson, J. Incandela, S. Koay, R. Rossin}
\affiliation{UCSB, Santa Barbara, CA 93106, USA}
\author{C. Hill}
\affiliation{University of Bristol, Bristol BS8 1TH, UK}

\begin{abstract}
We present three measurements of the top quark mass in the lepton plus jets channel with \lumi\ of data
using quantities with minimal dependence on the jet energy scale in the lepton plus jets channel at CDF. One measurement
uses the mean transverse decay length of \Bot-tagged jets (L2d) to determine the top mass, another uses the transverse momentum of the lepton (LepPt) to determine the top mass, and a third measurement uses both variables simultaneously.

Using the L2d variable we measure a top mass of \lxyRes, using the LepPt
variable we measure a top mass of \lepPtRes, and doing the combined measurement
using both variables we arrive at a top mass result of \combRes. Since some of
the systematic uncertainties are statistically limited, these results are
expected to improve significantly if more data is added at the Tevatron in the
future, or if the measurement is done at the LHC.

\end{abstract}

\maketitle

\thispagestyle{fancy}


\section{\label{sec:Intro}Introduction}

This analysis grew out of a measurement of the top mass using the transverse decay length of \Bot-jets in the lepton
plus jets channel with 695 \ipb\ of data \cite{bib:Lxy_PRD_rapid}.
Since the measurement relies almost exclusively on tracking, very different
event information is used compared to conventional mass measurements, and so
the result is expected to have small correlation with other measurements.
Further, the jet energy scale systematic, which is the largest uncertainty on
the world average top mass, should have a minimal effect on the results.
However, the statistical resolution on the top mass is poor
relative to more conventional, reconstruction based top mass measurements.
To improve statistics, in addition to incorporating
more than double the data of the previous measurement, it was decided that a
second variable should be included to further reduce the uncertainty.

The second variable, originally proposed by C. S. Hill et. al. \cite{bib:Lxy_PRD}, was the transverse momentum of the leptons
(transverse energy as measured in the
calorimeter for electrons), which are correlated to the top mass through the
momentum of the W bosons. Measuring the top mass using the lepton
transverse momentum has a similar statistical power to the the decay length
technique, and has the advantage of being almost completely uncorrelated
statistically. Since the distribution of
both these variables is approximately that of a decaying exponential over most
of their range, the shapes are largely
specified by the mean of the distributions. Thus, to keep the analysis simple, we use only the means and not the shapes of these distributions to evaluate the top mass. 

\section{\label{sec:Data}Data Sample \& Event Selection}

The data for this analysis are collected with an inclusive lepton trigger that requires
an electron (muon) with E$_T$ (P$_T$) $>$ 18 GeV. From this inclusive lepton dataset we select events offline
with a reconstructed isolated lepton with E$_T$ (P$_T$) $>$ 20 GeV. All leptons used are required to be isolated and
have a well resolved track in the central tracking chambers. 

The total missing transverse energy (MET) in the event is required to be
greater than 20 GeV, and a minimum of three jets must also be identified with
reconstructed transverse energies greater than 20 GeV. \Bot-jets are identified
(tagged) using the SecVtx algorithm \cite{bib:SecVtx}. This algorithm
identifies tracks that are displaced from the primary vertex and attempts to
reconstruct a secondary vertex from them. If the secondary vertex is well
resolved and has a sufficient transverse decay length from the primary vertex,
the jet is tagged as a \Bot. Note that this decay length is also directly used later to
measure the top mass. In order for the event to pass selection, at least one jet must be tagged as a \Bot\ for events with
four or more jets of \ET\ greater than 20 GeV, and at least two jets must be
tagged as a \Bot\ for events with exactly three jets of \ET\ greater than 20 GeV.

\section{\label{sect:cor} Event Composition and Corrections}

The \ttbar\ signal Monte Carlo is generated in Pythia using the CTEQ5L \cite{bib:cteq5l} parton distribution function. Since this analysis is sensitive to inaccuracies in event kinematics, the events are reweighted to match the more accurate CTEQ6M \cite{bib:cteq6m} parton distribution function, and to match the expected (mass dependent) gluon fusion fractions. Further corrections (a scale factor) are applied to compensate for Monte Carlo mismodeling of the average \Bot-jet decay length. 

\begin{figure*}
\centerline{
  \mbox{\includegraphics[height=2.5in]{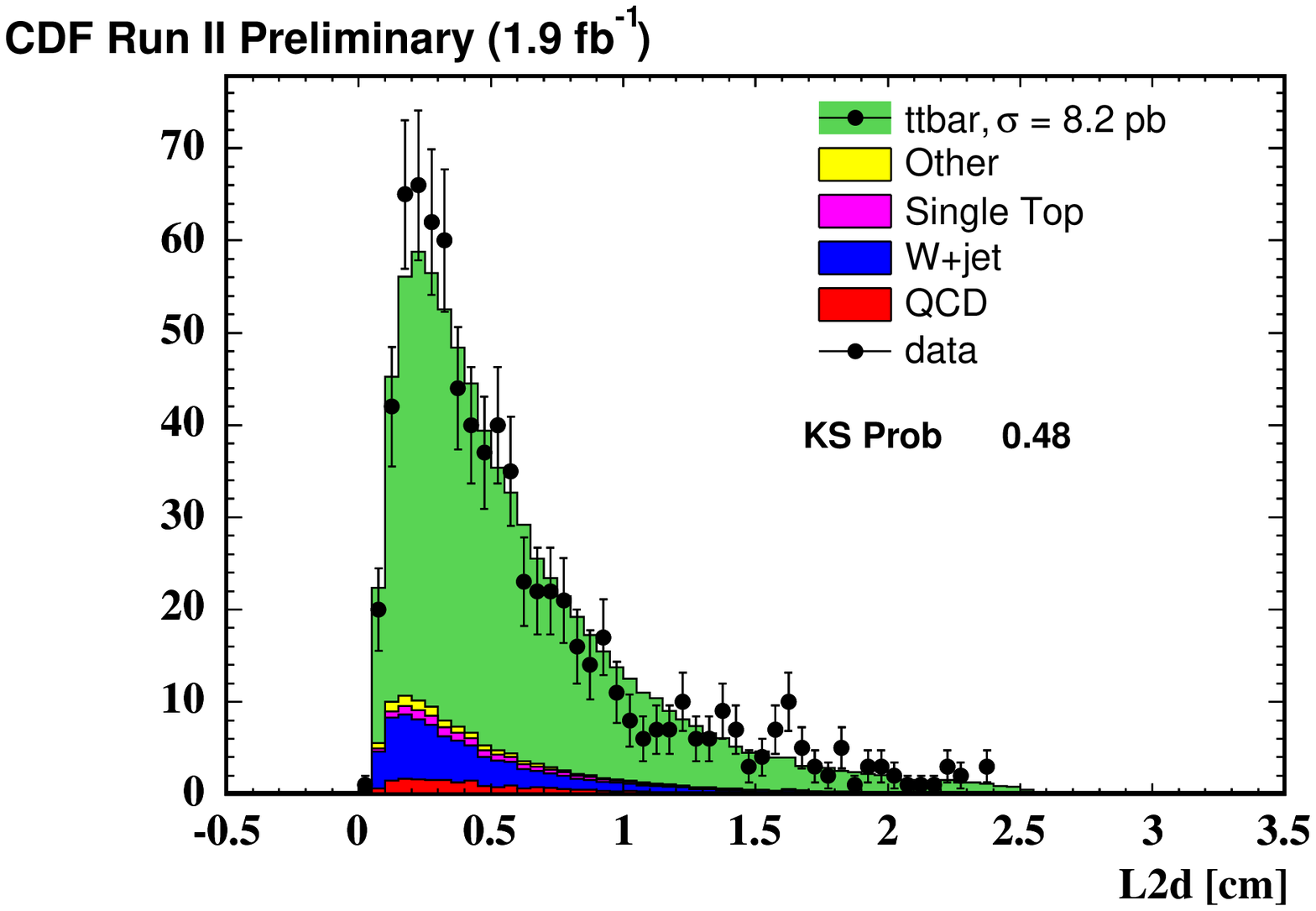}}
  \mbox{\includegraphics[height=2.5in]{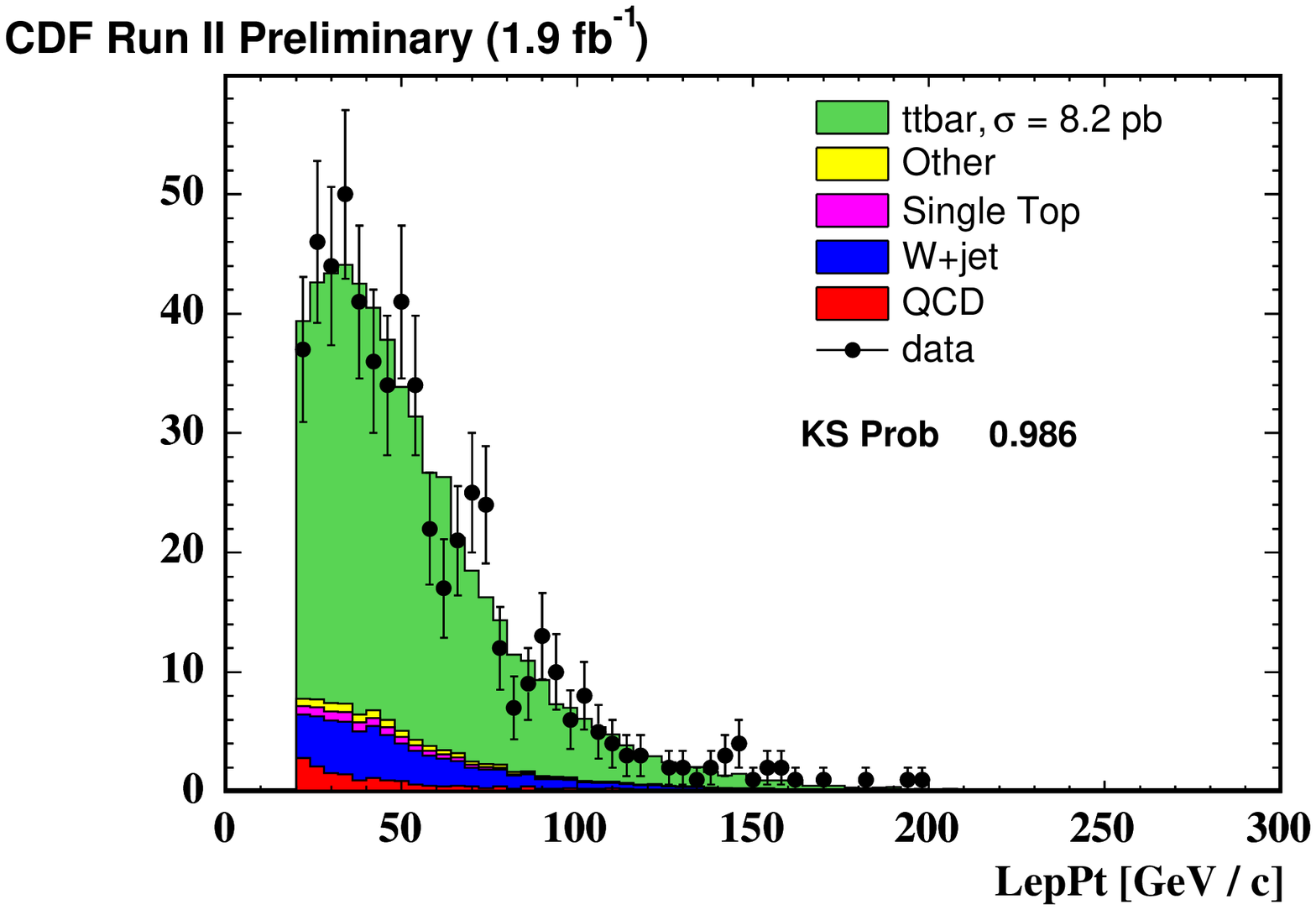}}
}
\caption{Left: Signal, background, and data for the L2d distribution using hypothesis top mass M=178 $GeV/c^2$. Right: Signal, background, and data for the LepPt distribution using hypothesis top mass M=173 $GeV/c^2$.}
\label{stacks}
\end{figure*}

Aside from the signal, the largest contributions come from W+jets and QCD events. The QCD background is evaluated from
data by altering the lepton selection criteria to make the events much more
likely to contain fake leptons, and the W+jets events are evaluated from ALPGEN Monte Carlo that is showered with Pythia. Expected signal and background distributions for the L2d and LepPt variables are shown with data in Figure~\ref{stacks}.

\section{\label{sec:Method} Method}

Pseudoexperiment events are drawn separately from signal and background
samples and combined according to the measured \ttbar\ cross section in the greater than three jet bins (8.2 pb). This is done separately for 24 hypothesis top mass values ranging from a top mass of $130\ GeV/c^2$ to a top mass of $220\ GeV/c^2$. For each pseudoexperiment the total number of events is fixed to the value observed in the data. The total number of background events is fluctuated within statistical and systematic uncertainties, and the rest of the events are considered to be signal.

\begin{figure*}
\centerline{
  \mbox{\includegraphics[height=2.5in]{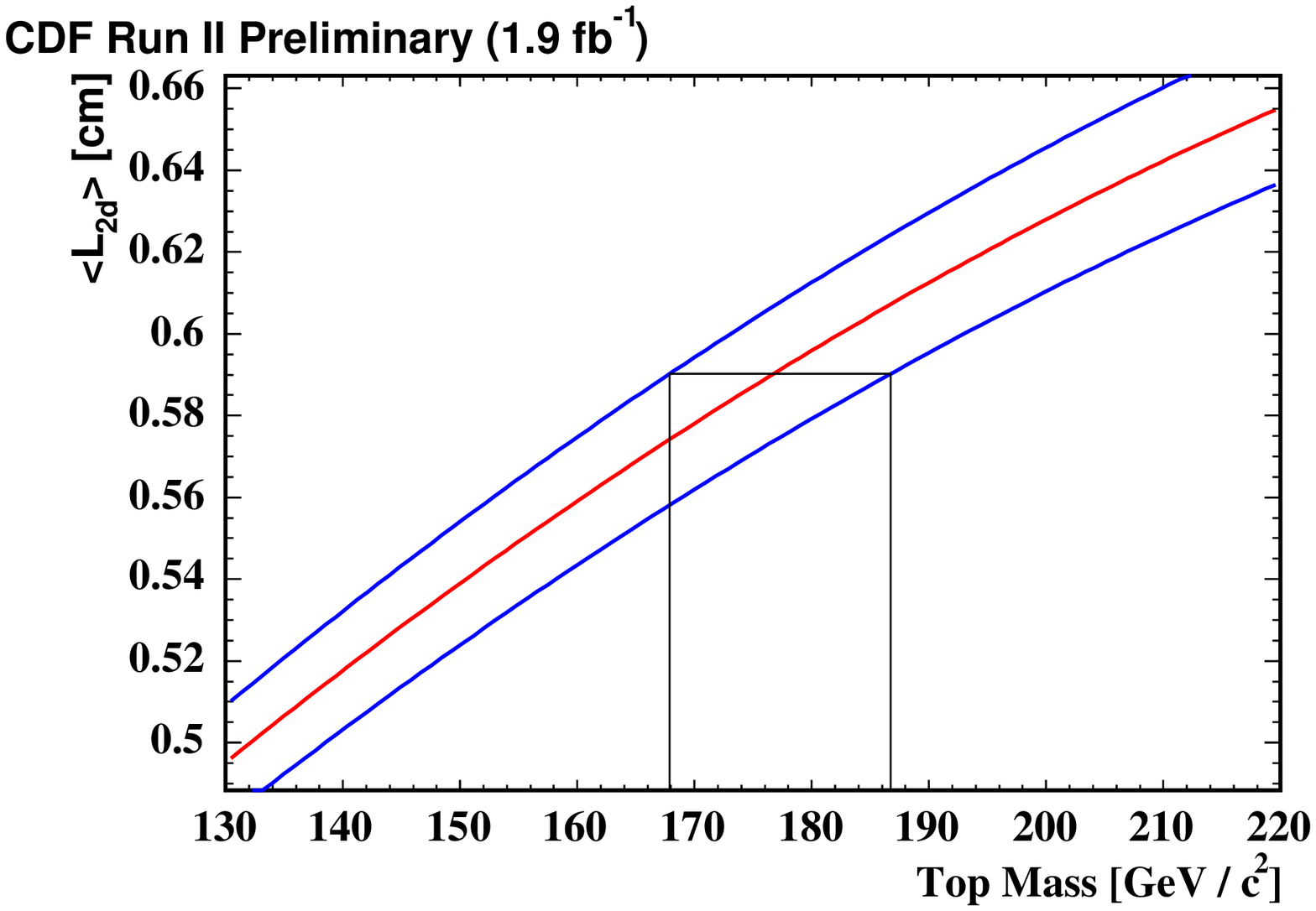}}
  \mbox{\includegraphics[height=2.5in]{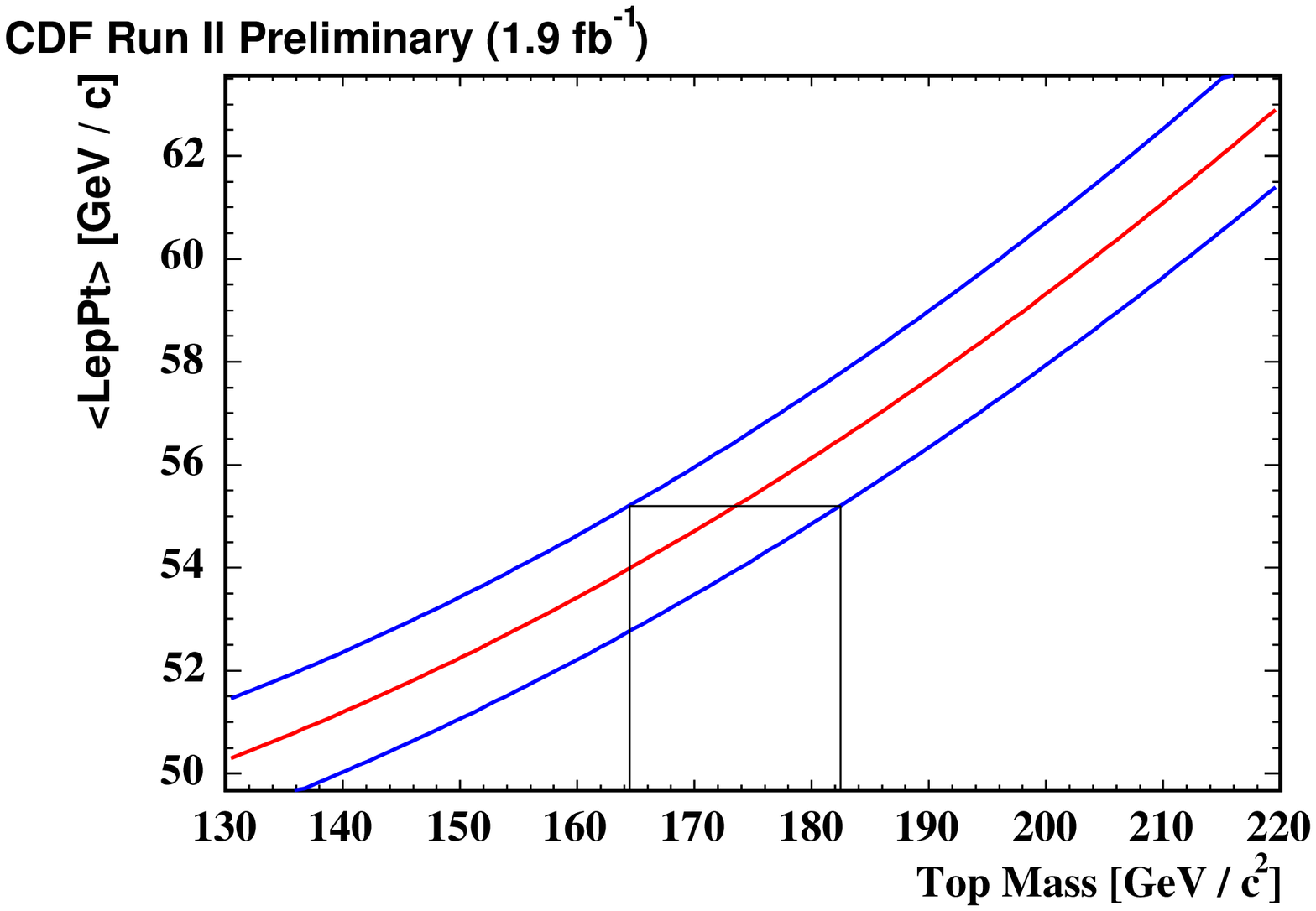}}
}
\caption{Left: Expected central values and one sigma confidence intervals of L2d mean results depending on top mass. Right: Expected central values and one sigma confidence intervals of LepPt mean results depending on top mass. The black lines show the plus and minus one sigma statistical uncertainties from data.}
\label{stat_results}
\end{figure*}

To evaluate the top mass results for each individual measurement (before L2d and
LepPt are combined), the means and RMS's of the pseudoexperiment results are determined and are
fit to quadratic polynomials as shown in Figure \ref{stat_results}. Given the mean L2d and LepPt in data, the corresponding
x-values of the central fit give us our expected mass, and the value of the
shifted fits give us our $\pm$ one sigma asymmetric statistical
uncertainties.

\section{\label{sec:Combination} Combination}

A joint top mass measurement using both the L2d and LepPt is also performed using pseudoexperiments. For each of the pseudoexperiments thrown for the L2d and LepPt measurements, the means are recorded. For a given top mass sample, these pseudoexperiments form an ellipse in the mean L2d versus mean LepPt plane. When results are measured in the data, they are compared to each of these hypothesis mass ellipses. The consistency of the mass hypothesis is evaluated based on the "distance" the data means are from the expected results (the ellipse center). This distance is evaluated from Equation~\ref{CorrEquation}:

\begin{equation}
D = \sqrt{ (\frac{\delta P_t}{{{\sigma}_P}_t})^2 + (\frac{\delta L_{2d}}{{{\sigma}_L}_{2d}})^2 }
\label{CorrEquation}
\end{equation}

\begin{figure}[tbp] \begin{center}
\includegraphics[height=2.5in, width=6.00in]{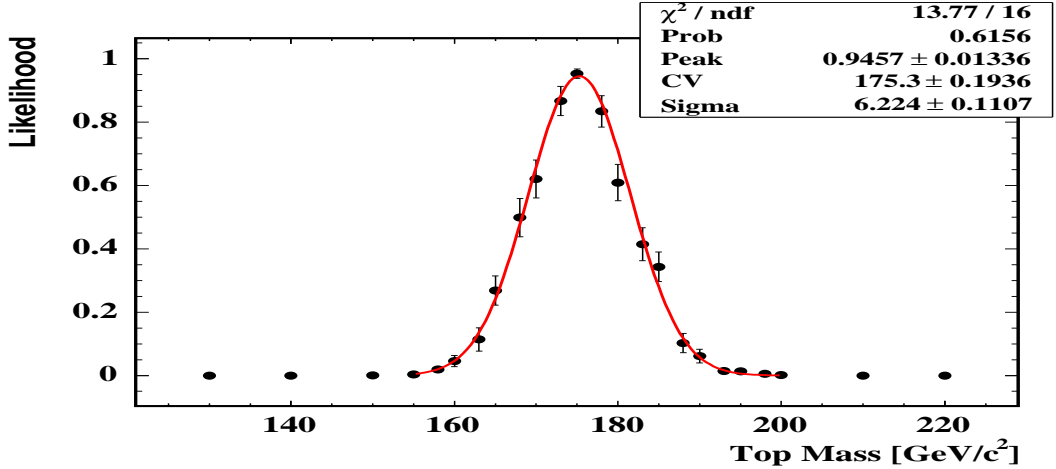}
        \caption{Likelihood fit results for data under the combined measurement.\label{likeli_fit}}
\end{center}\end{figure}

Here, $\delta P_t$ is the difference between the mean LepPt of the data and the
hypothesis value, and ${{\sigma}_P}_t$ is the size of the RMS of the hypothesis LepPt means from pseudoexperiments, etc.  If a given mass hypothesis represents the true value of the top mass, then the probability that a given pseudoexperiment will be at least as discrepant as the data is given by the fraction of pseudoexperiments with a larger value of $D$ than the data. These fractions are evaluated along with uncertainties (dictated by number of pseudoexperiments and finite Monte Carlo statistics) and are fit to a Gaussian. Results for the data are shown in Figure~\ref{likeli_fit}. This fit provides us with our mass result and our statistical uncertainty. These fits have been shown to be without bias and to properly produce Gaussian statistical uncertainties in nineteen Monte Carlo samples, ten of which were blinded in advance.

\section{Results}

Using \lumi\ of CDF data, we find 576 events passing our selection. From these events we measure a mean LepPt of \dataLepPt\, and a mean
L2d of \dataLxy\ (after application of the L2d scale factor, PDF, and gluon fusion reweightings, as explained previously). The associated top mass results and statistical uncertainties come out to \lxyStatRes\ for L2d and \lepPtStatRes\ for LepPt. Under the combined measurement the fit result shown in Figure~\ref{likeli_fit} returns us a top mass of \combStatRes.

\section{Systematic Uncertainties}

We evaluate the mean LepPt and L2d for data and compare to our background
estimations in the one and two jet bins as cross checks for our background
modeling. To be conservative, we take the larger of the shifts between the
expected and the observed results for the one and two jet bins as the
systematic uncertainty on our backgrounds. A number of factors are taken into
account in evaluating uncertainties on the signal. QCD radiation uncertainties
are evaluated using Pythia Monte Carlo samples generated with initial and final
state radiation simultaneously tuned up or down. The larger of the shifts
between the central value and the alternate radiation samples is taken as the
systematic uncertainty. For the LepPt measurement this comes out to a
surprisingly large number. This is because both the samples with more and less
QCD radiation end up having a larger mean LepPt than the nominal sample. To get
a handle on other uncertainties related to the generator, the top mass is
evaluated again using a Herwig Monte Carlo sample instead of Pythia. Using the
Herwig sample alters a number of properties, including how the QCD
fragmentation is performed, the QED radiation (which in Herwig is added in at
the leptonic W-decay vertices using the PHOTOS algorithm \cite{bib:photos}),
the transverse Fermi motion of the colliding partons, and spin correlations
between the top quarks. The full shift between the top mass results for the
Pythia and Herwig samples is taken as a conservative estimate of the
uncertainties due to these different generators. The parton distribution
function systematic is evaluated by reweighting to the twenty 90\% CTEQ6M \cite{bib:cteq6m}
eigenvector PDFs and adding the shifts in quadrature (allowing gluon fractions
to vary). Since these eigenvectors do not account for uncertainties on the
strong coupling constant, this extra uncertainty is determined by reweighting
to the CTEQ6A and CTEQ6B PDFs \cite{bib:cteq6ab}. These samples are similar to
the nominal CTEQ6M NLO PDF, but with altered values of the strong coupling
constant. 

As mentioned in~\ref{sect:cor}, the mean L2d results are corrected to account
for differences in the Monte Carlo modeling of the decay length compared to
data. The uncertainty on this correction (scale factor uncertainty) is
significant, but will improve with statistics. A similar uncertainty is
determined for the LepPt measurement due to uncertainty in the lepton
transverse momentum. This uncertainty is found by fitting the Z-mass peak in
data and Monte Carlo using electrons and muons separately. The observed shift
is not corrected for. Rather, to be conservative, the weighted average of the
full shift between the lepton types is taken as a systematic uncertainty. The
jet energy scale is the dominant uncertainty for many other top mass
measurements. In our case, the only possible way for the jet energy scale to
have any effect is in the way it changes event selection (based upon which jets
pass selection, and whether the MET passes the cut). It turns out that the
LepPt measurement is only minimally effected by the jet
energy scale, however the L2d measurement suffers a larger shift. This is
because tagged jets near the 20 GeV threshold have a significantly lower than normal
average decay length. At such low energies, the jet energy scale uncertainty is
entirely dominated by uncertainties in out-of-cone effects on the jet energy
scale. Improvements in the understanding of out-of-cone effects would help
reduce this systematic. This systematic can also be reduced at the cost of
statistics by cutting out low decay-length jets in the event selection. A
summary of the systematic uncertainties is shown in Table~\ref{syst_table}.

\begin{table}[th]
  \begin{center}
  \caption{Systematic Results}\label{syst_table}
  \begin{tabular}{|l|c|c|c|}
  \hline
  Systematic & L2d & LepPt & Combination \\
  \hline

  QCD Radiation & 0.9 & 2.3 & 1.5 \\
  PDFs & 0.3 & 0.6 & 0.5\\
  Generator & 0.7 & 1.2 & 0.6 \\     
  L2d Scale Factor & 2.9 & 0 & 1.4\\
  LepPt scale & 0 & 2.3 & 1.1\\
  Bkg Shape & 1.0 & 2.3 & 1.6\\
  Out of Cone JES & 1.0 & 0.3 & 0.6\\
  Total & 3.4 & 4.2 & 3.0 \\

  \hline
  \end{tabular}
  \end{center}
\end{table}

\section{Conclusion}

We have performed three measurements of the top quark mass using variables with
minimal correlation to the jet energy scale and combined them. Under an
integrated luminosity of \lumi\ we measure a top quark mass of \lxyRes\ using the decay
length method, \lepPtRes\ using the lepton transverse momentum, and \combRes\ in
combination.  If updated, the results of this method will improve, but will
continue to be limited by statistics for the rest of Run II. However, if this
analysis is done at the LHC statistics will no longer be an issue. Further,
since some of the dominant systematics are statistically limited, the results
of these techniques could well become competitive with conventional top mass
analyses, and due to their reduced correlation with conventional top
measurements they should help reduce the uncertainty on the world average top
mass in a combination.

\end{document}